\documentclass[onecolumn]{emulateapj}
\usepackage{amssymb,amsfonts,amsmath,amstext,amsgen,amsopn,amsxtra,indentfirst,lscape}
\shorttitle{Planetesimal Disk Microlensing}
\shortauthors{Heng \& Keeton}

\begin{document}

\title{Planetesimal Disk Microlensing}

\author{Kevin Heng\altaffilmark{1,2} \& Charles R. Keeton\altaffilmark{3,1}}

\altaffiltext{1}{Institute for Advanced Study, School of Natural Sciences, Einstein Drive, Princeton, NJ 08540, U.S.A.}
\altaffiltext{2}{Frank \& Peggy Taplin Member.  Email: heng@ias.edu}
\altaffiltext{3}{Department of Physics \& Astronomy, Rutgers University, 136 Frelinghuysen Road, Piscataway, NJ 08854, U.S.A.  Email: keeton@physics.rutgers.edu}

\begin{abstract}
Motivated by debris disk studies, we investigate the gravitational microlensing of background starlight by a planetesimal disk around a foreground star.  We use dynamical survival models to construct a plausible example of a planetesimal disk and study its microlensing properties using established ideas of microlensing by small bodies.  When a solar-type source star passes behind a planetesimal disk, the microlensing light curve may exhibit short-term, low-amplitude residuals caused by planetesimals several orders of magnitude below Earth mass.  The minimum planetesimal mass probed depends on the photometric sensitivity and the size of the source star, and is lower when the planetesimal lens is located closer to us.  Planetesimal lenses may be found more nearby than stellar lenses because the steepness of the planetesimal mass distribution changes how the microlensing signal depends on the lens/source distance ratio.  Microlensing searches for planetesimals require essentially continuous monitoring programs that are already feasible and can potentially set constraints on models of debris disks, the supposed extrasolar analogues of Kuiper belts.
\end{abstract}

\keywords{gravitational lensing --- methods: data analysis --- Kuiper belt --- planets and satellites: general --- debris disks}

\section{Introduction}
\label{sect:intro}

The advent of infrared (IR) observatories, such as {\it Spitzer}, has revealed a large number of old ($\gtrsim 10^8$ yr) disks with IR excesses, known as ``debris disks'' (Zuckerman 2001; Wyatt 2008).  It is generally believed that the excess IR emission emanates from dust produced by collisions between planetesimals in these gas-poor disks.  Despite efforts to link the IR emission, dust grains and planetesimal populations (Krivov et al.\ 2008), the properties of the planetesimals remain poorly understood.  Fundamentally, IR observations only probe the collisional cascade of dust grains and are mute on the mass distribution of the primordial planetesimals (Heng \& Tremaine 2009).  It is fair to say that no robust constraints have been set on the planetesimal population from studies of debris disks.  In fact, how to determine the masses of the largest planetesimals remains an open question (Wyatt \& Dent 2002).  To date, the only empirical constraints on planetesimals come from our own Kuiper Belt (Luu \& Jewitt 2002).  This provides incentive to develop alternative ways to detect planetesimals, in order to better understand the planetesimals themselves and also to enhance the scientific impact of planned surveys of debris disks (e.g., the {\it SEEDS} survey by {\it Subaru}).

Gravitational microlensing (e.g., Paczy\'{n}ski 1996) is a rapidly maturing field that offers such an alternative.  It is already an established way of detecting planets as is evident from recent discoveries (Bond et al.\ 2004; Udalski et al.\ 2005; Beaulieu et al.\ 2006; Gould et al.\ 2006; Bennett et al.\ 2008; Gaudi et al.\ 2008; Dong et al.\ 2009) and theoretical work (e.g., Mao \& Paczy\'{n}ski 1991; Gould \& Loeb 1992; Wambsganss 1997; Griest \& Safizadeh 1998; Gaudi et al.\ 1998).  Microlensing by small bodies has also been discussed (Bromley 1996; Agol 2002, 2003; di Stefano \& Scalzo 1999a); various possibilities include microlensing by systems with multiple planets (di Stefano \& Scalzo 1999b), wide-separation planets (Han et al.\ 2005), extrasolar moons (Bennett \& Rhie 2002) and Earth-like moons around ice giants (Han 2008).

Drawing on and extending some of these established ideas, we suggest that gravitational microlensing provides an attractive new way to study planetesimal disks that comprise a \emph{population} of planetesimals with a \emph{distribution} of masses.  While the instantaneous probability of microlensing is dominated by the most massive planetesimals in the disk, the total number of microlensing events during a sustained observational campaign is more sensitive to intermediate-mass planetesimals.  Microlensing can probe not only systems like the observed debris disks --- a subset of planetesimal disks that tend to be large and dynamically hot (Heng \& Tremaine 2009) --- but also disks that are dynamically cold and/or small.  By combining microlensing and IR observations of hot disks, one can potentially extract information about total disk masses and planetesimal mass distributions.\footnote{While absolute masses can be difficult to determine in microlensing (e.g., Gaudi 2002; Han et al.\ 2005), the mass {\it ratios} between bodies that contribute to a microlensing light curve can be measured well.}  With good constraints on planetesimal mass distributions at the low- and high-mass ends, one can infer disk ages (Pan \& Sari 2005).  In general, detecting planetesimal disk microlensing events will provide valuable new constraints on models of debris disks.

The novelty of our approach stems from uniting dynamics and microlensing.  Since the formation of planets and planetesimal belts from first principles is fraught with uncertainties (Goldreich et al.\ 2004), we use a simple survival model to construct a plausible planetesimal disk in \S\ref{sect:disk}; we focus on one example of a large, hot disk.  In \S\ref{sect:microlensing}, we tap into some ideas from the microlensing literature, use them to discuss the basic phenomenology of microlensing by planetesimal disks, and give simple but realistic estimates of the occurrence of disk microlensing events.  We argue that long-term monitoring for low-magnification events will inform us about the properties of planetesimals that are otherwise invisible to conventional methods of detection.  The implications of our results and opportunities for future work are discussed in \S\ref{sect:discussion}.

\section{Disk Properties}
\label{sect:disk}

\subsection{Disk Parameters from a Survival Model}
\label{subsect:dynamics}

A planetesimal disk is characterized by its age ($t_{\rm age}$), mass ($M_{\rm disk}$) and semi-major axis ($a$), and is composed of planetesimals with a range of masses $m$ and radial velocity dispersions $\sigma_r$.  To have an age of $t_{\rm age}$, a disk must have survived all dynamical processes during that time (Heng \& Tremaine 2009).  Planetesimal disks may be broadly separated into those for which planetesimal orbits do or do not cross within their lifetime, respectively termed ``hot'' or ``cold'' disks.  For the purpose of discussion, we adopt the following disk parameters:
\begin{equation}
t_{\rm age} = 10^8 \mbox{ yr},\quad
M_{\rm disk} = 10 M_\earth,\quad
a = 10 \mbox{ AU}.
\label{eq:parameters}
\end{equation}
Our assumption for $a$ is plausible because dozens of debris disks have been found with $1 \lesssim a \lesssim 100$ AU (see Figure 7 of Wyatt 2008).  The assumed age is also a factor $\sim 10$ longer than conceivable time scales for protoplanetary disks to disperse their gas (Hillenbrand 2008; see also Figure 2 of Wyatt 2008 and references therein).  In addition, we assume the internal density of the planetesimals to be $\rho_p = 3$ g cm$^{-3}$ and the mass of the parent star to be $M_\star = M_\sun$.

The disk mass can be expressed as
\begin{equation}
M_{\rm disk} = 2\pi \int^{a_{\rm out}}_{a_{\rm in}} ~\Sigma\left(a^\prime\right) ~a^\prime ~da^\prime,
\label{eq:diskmass}
\end{equation}
where $\Sigma$ is the mass surface density, and the inner and outer disk radii are $a_{\rm in} = a/\eta$ and $a_{\rm out} = a \eta$.  We adopt
\begin{equation}
\Sigma(a^\prime) \propto \left(\frac{a^\prime}{a}\right)^{-3/2} ,
\label{eq:Sigma}
\end{equation}
as is assumed for models of the minimum mass solar nebula (Weidenschilling 1977).  Such a scaling also implies that the Toomre parameter, $Q=\sigma_r \Omega/ \pi G \Sigma$, is independent of the semi-major axis, where $\Omega = \sqrt{GM_\star/a^3}$ is the orbital angular velocity.  Equation (\ref{eq:diskmass}) can be rewritten as
\begin{equation}
M_{\rm disk} = f_m \pi a^2 \Sigma\left(a\right),
\end{equation}
and setting $f_m = 1$ then implies $\eta \approx 1.28$.  In other words, our fiducial disk has a radial width of $a_{\rm out} - a_{\rm in} \approx 0.5 a$, and an area $A_{\rm disk} = \pi(a_{\rm out}^2 - a_{\rm in}^2) \approx \pi a^2$.

The maximum allowed planetesimal mass and radial velocity dispersion are mainly determined by the requirements that the timescale for gravitational scattering exceeds $t_{\rm age}$ and that the disk is thin.  The scattering condition reads:
\begin{equation}
t_g = \left[\frac{d\ln{\left(e^2_0\right)}}{dt} \right]^{-1} \gtrsim t_{\rm age},
\label{eq:diskcon1}
\end{equation}
where $e_0$ is the root mean square eccentricity of the planetesimal orbits.  The left-hand side can be evaluated using equation (3.29) of Stewart \& Ida (2000), where we take the root mean square inclination of the planetesimal orbits to be $i_0 = e_0/2$.  

Taking $h$ to be the root mean square disk height, the thin disk condition reads:
\begin{equation}
h \lesssim 0.35 f_e a,
\label{eq:diskcon2}
\end{equation}
which can be interpreted as the condition that most of the planetesimals are bound or that the radial velocity dispersion is less than the circular speed.  The preceding expression is equivalent to $e_0 \lesssim f_e$ or $\sigma_r \lesssim f_e \Omega a/\sqrt{2}$; we adopt $f_e = 0.5$ (Heng \& Tremaine 2009).

Equations (\ref{eq:diskcon1}) and (\ref{eq:diskcon2}) together yield:
\begin{equation}
\begin{split}
&m \lesssim 2M_\earth,\\
&\sigma_r \lesssim 3.4 \mbox{ km s}^{-1} ~~\left(\approx 0.4 ~a\Omega \right).\\
\end{split}
\end{equation}
For comparison, the circular speed at $a=10$ AU is $a\Omega \approx 9.4$ km s$^{-1}$, while the typical bulk velocities of stars near the Galactic bulge\footnote{For illustration we consider source stars near the bulge because they are common targets for microlensing campaigns, but as we shall see it will be possible to consider other source locations without an appreciable loss of signal.} are $\sim 100$ km s$^{-1}$.  Thus, as a first approximation we may consider that microlensing is driven by the bulk motions of the lens and source, and neglect the motions of the planetesimals within the disk.

For our fiducial disk parameters, Toomre stability is trivially fulfilled.  Note that for $m \approx 2M_\earth$ and $\sigma_r \approx 0.4 ~a\Omega$, the Safronov number is $\Theta \gg 1$.  Therefore, $t^{-1}_{\rm age} \gtrsim t^{-1}_g \propto \Theta^2$ is a stronger condition than $ t^{-1}_{\rm age} \gtrsim t^{-1}_c \propto \Theta$, where $t_c$ is the collision time.

Further details of dynamical survival models for long-lived planetesimal disks are described in Heng \& Tremaine (2009).

\subsection{Planetesimal Population}
\label{subsect:popn}

We assume the planetesimals have a mass distribution such that the number of planetesimals between mass $m$ and $m+dm$ is
\begin{equation}
\frac{dN}{dm}\ dm = B m^{-\alpha}\ dm.
\label{eq:size_dist}
\end{equation}
Alternatively, one can instead write the size distribution, $dN/dr \propto r^{-q}$, where $q = 3\alpha - 2$ and $r = (3m/4\pi\rho_p)^{1/3}$.  Dohnanyi (1969) showed that $\alpha = 11/6$ ($q=7/2$) in a steady-state system.  Pan \& Sari (2005) rederived and generalized the results of Dohnanyi (1969), allowing for collisions to be inefficient (i.e., the kinetic energy of the bullet does not entirely go into breaking up the target); they inferred that $23/8 < q < 22/7$ or $13/8 < \alpha < 12/7$.  While the mass function may not be a single power law over the full range of interesting masses (as in our Kuiper Belt; Bernstein et al.\ 2004), in this pilot study we use a single power law.  If we restrict ourselves to $\alpha < 2$, and we assume that the smallest planetesimal mass in the distribution is much smaller than the largest mass $m_{\rm L}$, then the normalization factor is
\begin{equation}
B \approx \left( 2 - \alpha \right) M_{\rm disk} m^{\alpha-2}_{\rm L} .
\end{equation}

For the purpose of microlensing, one needs to consider the \emph{projected} planetesimal number density per unit mass, $dn_{\rm proj}/dm$, such that
\begin{equation}
\frac{dN}{dm} = 2\pi \int^{a_{\rm out}}_{a_{\rm in}} ~\frac{dn_{\rm proj}}{dm} ~a^\prime ~da^\prime,
\end{equation}
from which it follows that
\begin{equation}
\frac{dn_{\rm proj}}{dm} = \frac{Bm^{-\alpha}}{\pi a^2} \left(\frac{a^\prime}{a}\right)^{-3/2}.
\end{equation}

\section{Disk Microlensing}
\label{sect:microlensing}

\subsection{Basic Picture}
\label{subsect:basic}

\begin{figure}
\begin{center}
\includegraphics[width=4.5in]{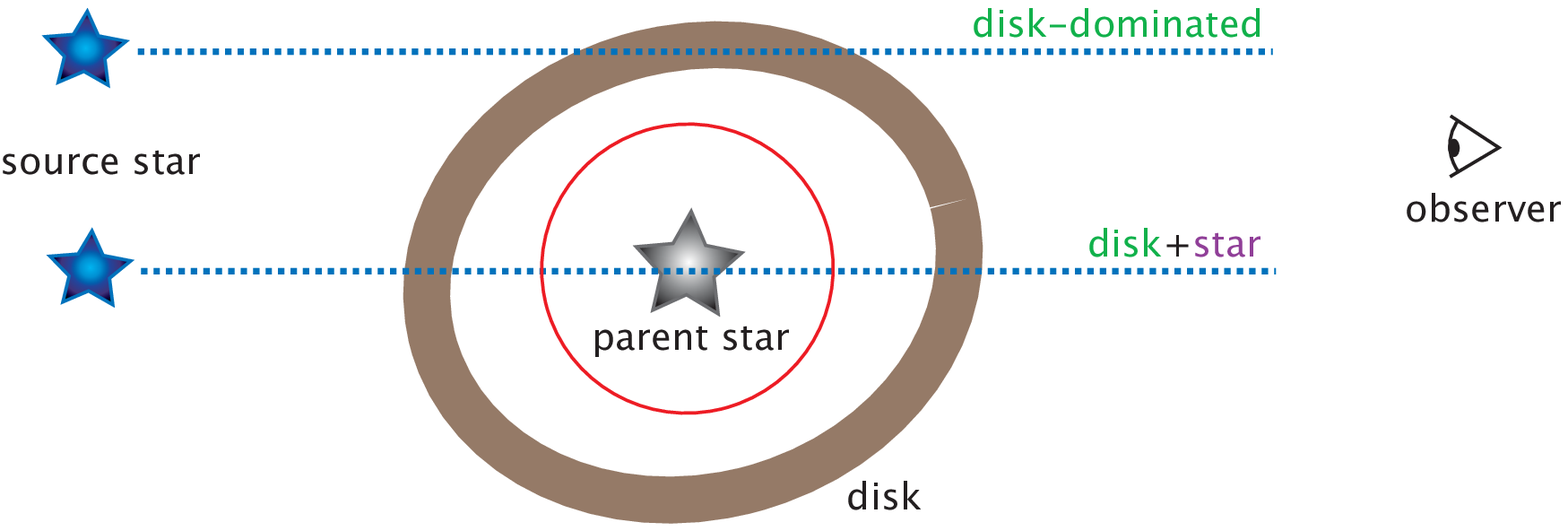}
\includegraphics[width=4in]{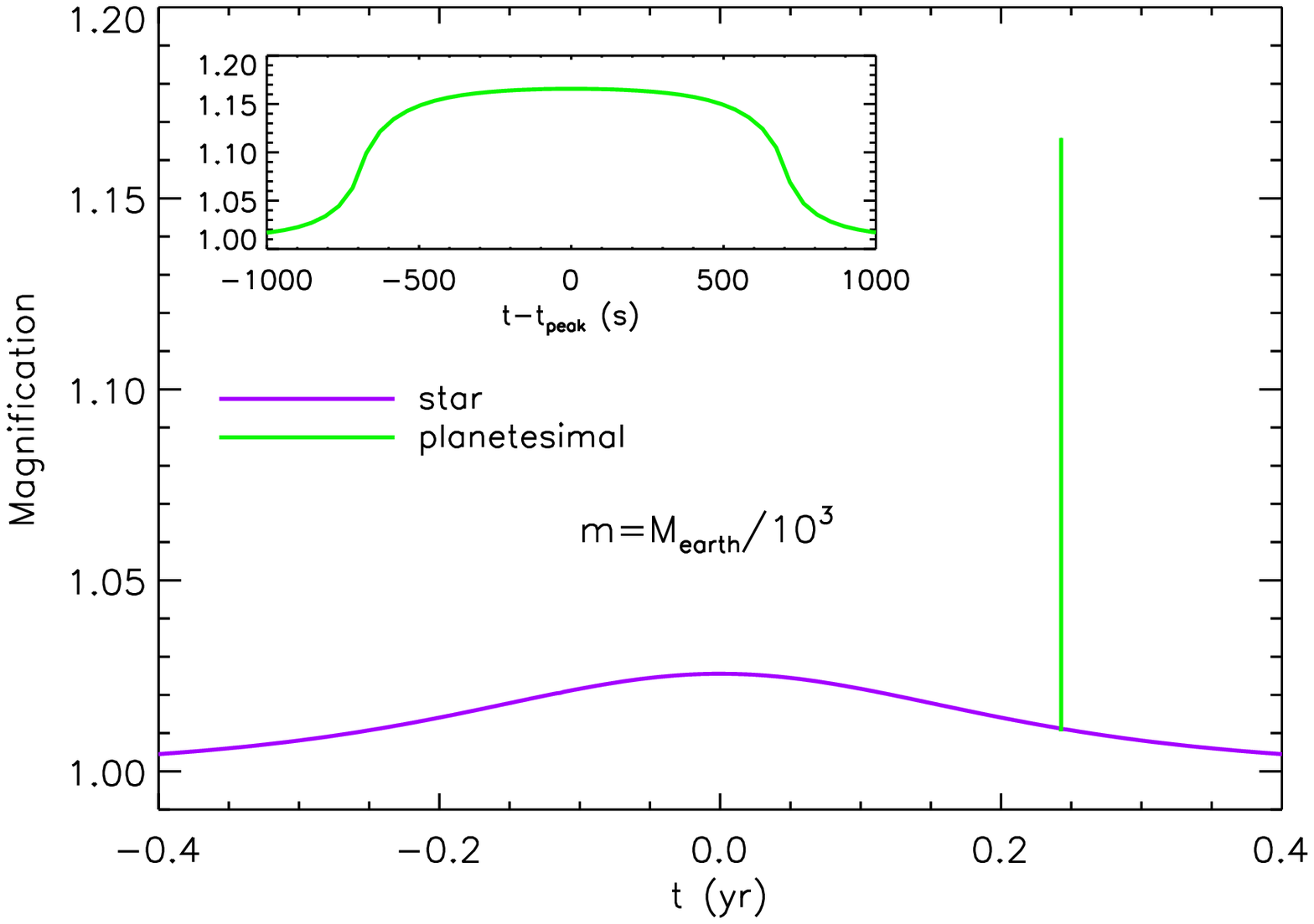}
\includegraphics[width=4in]{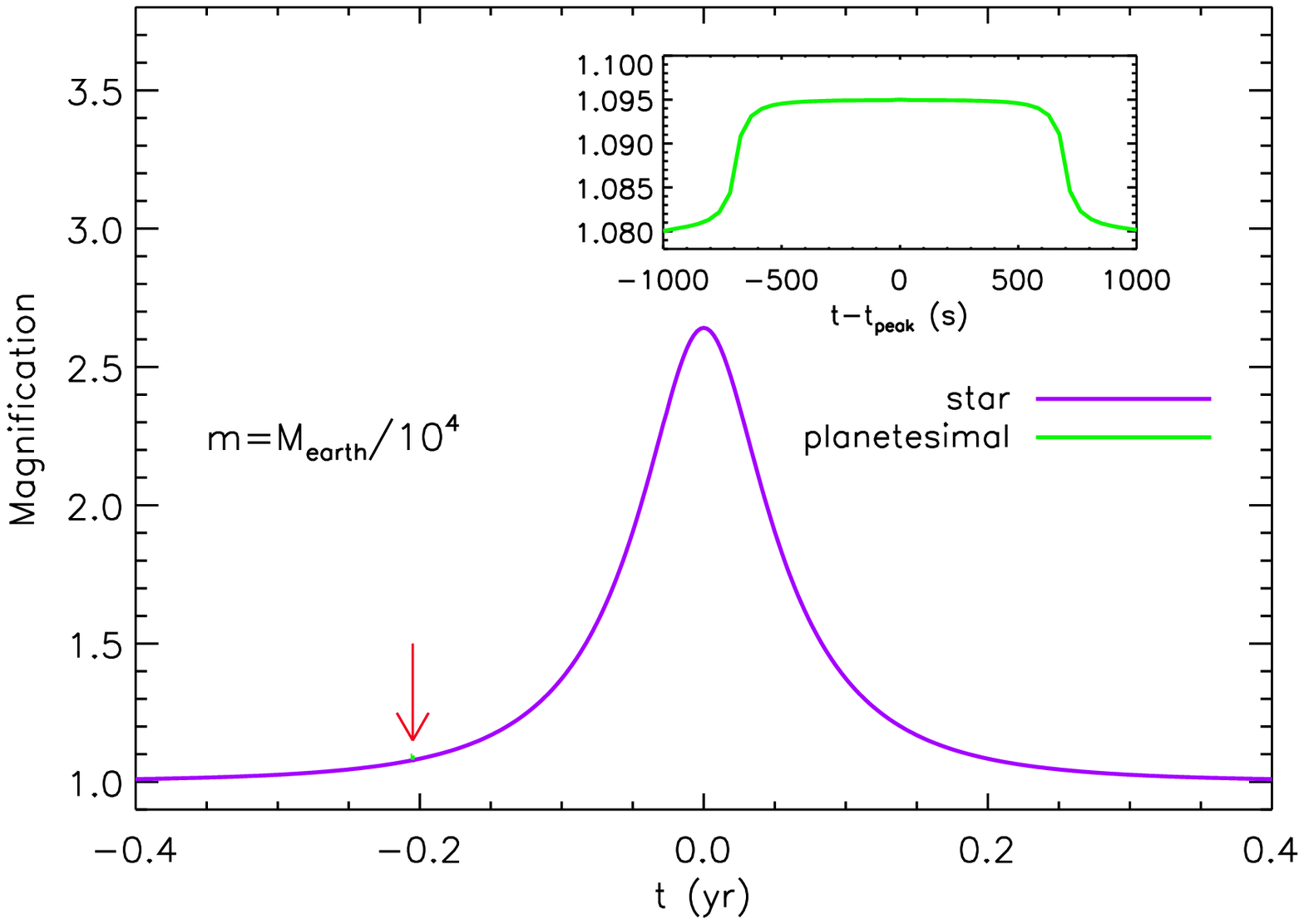}
\end{center}
\caption{The top panel shows a schematic diagram (not to scale) of a source star passing behind a fiducial planetesimal disk with $M_{\rm disk} = 10~M_\earth$, $a = 10$ AU, $D_s = 8$ kpc, $f_l = 0.1$, $m_{\rm L} = 2~M_\earth$ and $\alpha = 11/6$.  The thin red circle indicates the Einstein radius of the parent star (of mass $M_\sun$; $R_{\rm E\star} = 1.2$ AU), while the thick brown ring represents the planetesimal disk.  The dotted lines show sample source trajectories that lead to the light curves shown in the middle and bottom panels.  In the middle panel, microlensing by the star is weak ($\delta \sim 1\%$) compared to that by the planetesimal ($\delta \sim 10\%$).  In the bottom panel, the star dominates the microlensing lightcurve, while the planetesimal contributes a $\delta \sim 1\%$ residual.  For illustration, we have assumed $v_\perp=100$ km s$^{-1}$ for the purpose of computing $t$, the time of passage.  The insets zoom in on the light curves associated with the planetesimals; $t_{\rm peak}$ is the time at which the planetesimal microlensing events peak.  Note that the light curves are from full numerical calculations that do not involve the assumptions of isolation and no shear.  While this schematic serves to illustrate the basic idea of planetesimal disk microlensing, the brightness of the lens star for nearby events renders, for example, a 10\% deviation into a $\sim 0.1\%$ deviation for $f_l=0.1$ (see \S\ref{subsect:detectability}).}
\label{fig:cartoon}
\end{figure}

Figure~\ref{fig:cartoon} shows examples of a source star passing behind a fiducial disk.  If the source star passes close (in projection) to the parent star we get a classic stellar microlensing event, and if it also passes close to a planetesimal we can get a short secondary feature in the light curve.  This phenomenon is familiar from planetary microlensing (Mao \& Paczy\'{n}ski 1991; Gould \& Loeb 1992; Han et al.\ 2005); the main difference with planetesimal disk microlensing is that the masses are smaller and we must explicitly consider a significant \emph{population} of planetesimals.

Qualitatively, there are three limiting regimes of microlensing involving a planetesimal disk: (1) the source star passes directly behind a planetesimal that is ``far'' from its parent star; (2) the source passes directly behind a planetesimal that is ``near'' its parent star; or (3) the source star passes directly behind the parent star in such a way that the light curve is still sensitive to the presence of the planetesimal disk.  In the first two cases, ``far'' and ``near'' are defined with respect to the Einstein radius of the parent star,
\begin{equation}
R_{\rm E\star} = \frac{2}{c} \sqrt{G M \tilde{D}} = 4.0 \mbox{ AU} ~\left[\frac{f_l \left(1-f_l\right)}{0.25} \frac{D_s}{8 \mbox{ kpc}} \frac{M}{M_\sun} \right]^{1/2},
\end{equation}
where $\tilde{D} \equiv D_l D_{ls}/D_s$ given the distance to the lens ($D_l$), the distance to the source ($D_s$), and the distance from the lens to the source ($D_{ls} = D_s - D_l$).  It is convenient to define the lens/source distance ratio
\begin{equation}
f_l \equiv \frac{D_l}{D_s},
\end{equation}
and then write $\tilde{D} = f_l (1-f_l) D_s$.  The distinction between the ``far'' and ``near'' regimes arises because the star creates a tidal shear $\gamma = (R_{\rm E\star}/d)^2$ at a projected distance $d$, which enhances the cross section for microlensing by $\sim \gamma^2$.  In the ``far'' regime, we can neglect the influence of the parent star on microlensing by the planetesimal, and the system can be regarded as a wide-separation binary (di Stefano \& Scalzo 1999a; Han et al.\ 2005).  By contrast, in the ``near'' regime the microlensing signal is significantly affected by ``planetesimal caustics'' (i.e., analogues of planetary caustics; see Schneider \& Weiss 1986 for a full discussion of binary point-mass lens systems).  Finally, the third regime corresponds to a high-magnification event in which the light curve may be perturbed by secondary caustics that can be used to detect the presence of planets (Wambsganss 1997; Gaudi et al.\ 1998; Griest \& Safizadeh 1998), or even planetesimals.

The spatial scale for microlensing by a planetesimal is set by its Einstein radius,
\begin{equation}
R_{\rm E} = \frac{2}{c} \sqrt{G m \tilde{D}} \approx 10^{10} \mbox{ cm} ~\left(\frac{m}{0.01 M_\earth} \frac{\tilde{D}}{2 \mbox{ kpc}} \right)^{1/2}.
\end{equation}
For comparison, a planetesimal of mass $m = 0.01\,M_\earth$ and density $\rho_p = 3$ g cm$^{-3}$ has a physical size of $r \sim 10^8$ cm, implying that it can be treated as a point lens; we return to this issue in Figure \ref{fig:minmass}.  Note that since the Einstein radius scales as $R_{\rm E} \propto m^{1/2}$ while the physical size scales as $r \propto m^{1/3}$, there is some mass threshold below which the physical size is comparable to or larger than the Einstein radius.  When the two are comparable, it may be possible to infer the size from combined microlensing and occultation measurements (Bromley 1996; Agol 2002).

A basic time scale for microlensing is the time it takes to cross the Einstein diameter,
\begin{equation}
t_{\rm E,d} = \frac{2 R_{\rm E}}{v_\perp} \approx 35 \mbox{ min} ~\left(\frac{m}{0.01 M_\earth} \frac{\tilde{D}}{2 \mbox{ kpc}} \right)^{1/2} \left(\frac{v_\perp}{100 \mbox{ km s}^{-1}} \right)^{-1},
\label{eq:tein}
\end{equation}
where $v_\perp$ is the transverse relative velocity of the lens and source.  As discussed below, the actual duration of a microlensing event may be longer than $t_{\rm E,d}$ because of finite source effects.  In particular, as the lens mass decreases and $t_{\rm E,d}$ becomes shorter, finite source effects set a floor on the event duration given by the time it takes the lens to cross the diameter of the source star,
\begin{equation}
t_{\rm \star,d} = \frac{2 f_l R_\star}{v_\perp} \approx 116 \mbox{ min} ~\left(\frac{f_l}{0.5}\frac{R_\star}{R_\odot}\right) \left(\frac{v_\perp}{100 \mbox{ km s}^{-1}} \right)^{-1} ,
\label{eq:tsrc}
\end{equation}
where the factor of $f_l$ handles the projection of the source size into the lens plane.  The last time scale of interest is the time taken to cross the full planetesimal disk,
\begin{equation}
t_{\rm disk} \approx 1 \mbox{ yr} ~\left(\frac{a}{10\mbox{ AU}}\right) \left(\frac{v_\perp}{100\mbox{ km s}^{-1}}\right)^{-1}.
\label{eq:diskcrossing}
\end{equation}

\subsection{Figures of Merit}
\label{subsect:fom_general}

While some authors have used detailed simulations of microlensing campaigns to forecast the detection of planets (e.g., Han et al.\ 2005), we elect to use simple but reasonable figures of merit to give initial estimates of the occurrence of planetesimal disk microlensing.  Wherever possible, we compare our results with those in the literature.  We ignore factors of order unity related to the planetesimal disk geometry and assume a face-on disk.  In this paper, we make two assumptions: (1) the planetesimals are far enough from the parent star that we can neglect shear; (2) the planetesimals are effectively isolated from one another, so the light curve is influenced by only one planetesimal at a time.  The assumption of isolation can be verified \emph{a posteriori} by checking that the microlensing optical depth is low.  Relaxing these assumptions will provide interesting opportunities for follow-up work, as we discuss in \S\ref{sect:discussion}.

Microlensing increases the observed flux of the source star by the time-dependent magnification factor ${\cal A}(t)$ (Paczy\'{n}ski 1986).  In general, the total observed flux may include contributions not only from the amplified source star but also from the lens star and any other stars along the line of sight that are projected within the same resolution element:
\begin{equation}
  F_{\rm obs}(t) = {\cal A}(t) F_{\rm source} + F_{\rm lens} + F_{\rm other}\,.
\label{eq:fobs}
\end{equation}
We set $F_{\rm lens} = F_{\rm others} = 0$ in the present analysis for clarity and note that the effects of stellar blending (di Stefano \& Esin 1995; Han et al. 2006) must be kept in mind when considering prospects for detecting low-amplitude microlensing events (see \S\ref{subsect:detectability}).  In the absence of blending, the maximum fractional change in the flux is
\begin{equation}
  \delta \equiv \max_t\left[\frac{F_{\rm obs}(t)}{F_{\rm source}}-1\right]
  = \max\left[{\cal A}\left(t\right)\right] - 1.
\end{equation}
For the purpose of gaining a general understanding of planetesimal disk microlensing, we consider a microlensing event to be detectable if $\delta$ equals or exceeds some detection threshold $\delta_{\rm det}$.

A second factor that may be relevant for detection is the duration of the event.  Because of finite cadences and a desire to have more than one point on the light curve during a planetesimal disk microlensing event, there may be some minimum event duration $t_{\rm min}$ that can realistically be detected, allowing for the definition of two distinct regimes:
\begin{equation}
\begin{split}
t_{\rm \star,d} \ge t_{\rm min}: \mbox{ ``amplitude-limited''},\\
t_{\rm \star,d} < t_{\rm min}: \mbox{ ``duration-limited''},\\
\end{split}
\end{equation}
since the shortest possible event duration is given by the source crossing time (equation [\ref{eq:tsrc}]).  The duration-limited regime occurs when
\begin{equation}
  f_l < \frac{v_\perp t_{\rm min}}{2 R_\star}
  = 0.043 \left(\frac{t_{\rm min}}{10\mbox{ min}}\right)
    \left(\frac{v_\perp}{100\mbox{ km s}^{-1}}\right)
    \left(\frac{R_\star}{R_\odot}\right)^{-1} .
\label{eq:duration-limited}
\end{equation}
We have chosen the nominal duration threshold $t_{\rm min} = 10$ min based on cadences that can be achieved today (e.g., Dong et al.\ 2009) and those planned for next-generation microlensing campaigns (Bennett et al.\ 2009; Gaudi et al.\ 2009), but the result can be trivially rescaled.  An added complication might be to impose a longer duration threshold for lower-amplitude events, perhaps to conserve total signal-to-noise.

In the amplitude-limited regime, planetesimal disk microlensing events have durations equal to or exceeding $t_{\rm min}$ by definition, so we do not need to factor the duration threshold into our calculations.  In the duration-limited regime, some fraction of events (associated with low-mass planetesimals) will have durations shorter than $t_{\rm min}$, implying that: their chances of detection will be reduced due to finite sampling; and even if they are detected the light curve will not be temporally resolved, leading to questions regarding whether they are true detections (see \S\ref{subsect:detectability}).  The exclusion of these low-mass events will reduce the total expected number of detected events associated with a given planetesimal disk.  We do not explicitly consider the duration-limited regime in our calculations, because it depends on the details of a particular microlensing survey and on the velocity distribution of source stars.  Nevertheless, we shall emphasize the circumstances under which our conclusions will be affected by a minimum detectable duration.

Specifying the condition $\delta \ge \delta_{\rm det}$ is equivalent to requiring $b \ge b_\phi$, where $b$ is the impact parameter of the source star relative to the lens.  As a zeroth order estimate, we expect $b_\phi$ to be comparable to the Einstein radius $R_{\rm E}$.  It may be possible, however, to have a detectable perturbation even when $b_\phi$ exceeds $R_{\rm E}$.  To allow for this possibility, we write the general form
\begin{equation}
b_\phi = \phi R_{\rm E},
\label{eq:boost}
\end{equation}
and call $\phi$ a ``boost factor.''  Such a boost factor has been described by di Stefano \& Scalzo (1999a), albeit with no name.

A standard figure of merit is the optical depth, which gives the probability that microlensing is detectable at any given instant in time.  In the limit that microlensing shear by the parent star is unimportant, each planetesimal has a ``circle of influence'' with area $\pi b_\phi^2$ (equivalent to the  ``lensing regions'' described by di Stefano \& Scalzo 1999a), and the optical depth is the fraction of the projected area of the disk ($A_{\rm disk}$) that is covered by the circles of influence:
\begin{equation}
\begin{split}
\tau &= \int \left(2\pi \int^{a_{\rm out}}_{a_{\rm in}} ~\frac{dn_{\rm proj}}{dm} ~a^\prime ~da^\prime \right) \frac{\pi b^2_\phi}{A_{\rm disk}} ~dm\\
&= \int \frac{\pi b^2_\phi}{A_{\rm disk}} ~\frac{dN}{dm} ~dm\\
&\approx \frac{4 G \tilde{D}}{ \left(a c\right)^2} ~B \int \phi^2 ~m^{1-\alpha} ~dm.\\
\end{split}
\label{eq:tau}
\end{equation}
We have neglected a factor of order unity related to disk geometry in equation (\ref{eq:tau}).  The limits of integration for the mass integral are discussed below.

A second --- and perhaps more interesting --- figure of merit is the expected number of microlensing events in one disk crossing (equation [\ref{eq:diskcrossing}]).  Conceptually, this is given by the number of planetesimals in a band of width $2b_\phi$ centered on the chord traced by the source.  If we assume the source crosses the full diameter of the face-on disk, we have:
\begin{equation}
\begin{split}
{\cal N} &= \int \left( 2 \int^{a_{\rm out}}_{a_{\rm in}} ~\frac{dn_{\rm proj}}{dm} ~da^\prime \right) 2 b_\phi ~dm\\
&\approx \int \frac{2 b_\phi}{\pi a}~\frac{dN}{dm}~dm\\
&= \frac{4 \sqrt{G\tilde{D}}}{\pi a c} ~B \int \phi ~m^{1/2-\alpha} ~dm,
\end{split}
\label{eq:events}
\end{equation}
The leading factor of 2 in the first expression enters because the source crosses the planetesimal disk during both ingress and egress.  The number of events will be reduced relative to this estimate by the fraction of the disk diameter actually crossed by the source.  (Note that while the event {\it rate} depends on the relative velocities of the lens and source, the number of events in one disk crossing does not.)

We can combine ${\cal N}$ with the number of disk crossings to estimate the total number of disk microlensing events that will occur during a sustained survey.  Consider a survey that targets a number $N_\star$ of source stars for some period $t_{\rm survey}$, waiting for lenses pass in front (see \S\ref{subsect:strategies} for other strategies).  For simplicity, we assume all the source stars are at the same distance $D_s$.  Conceptually, we picture each disk tracing out a band of width $2 a_{\rm out}$ and length $v_\perp t_{\rm survey}$.  We combine that with the three-dimensional number density of lens stars, $n_l$, to compute the fraction of the sky covered by the bands, factor in the number of source stars to obtain the number of disk crossings, and then multiply by ${\cal N}$ to estimate the total number of planetesimal microlensing events during the survey
\begin{equation}
N_{\rm survey} = \int_{0}^{D_s} 2 a_{\rm out} v_\perp t_{\rm surv} n_l N_\star {\cal N}~dD_l.
\label{eq:Nsurv}
\end{equation}
Of particular interest is the scaling with $f_l$,
\begin{equation}
\frac{dN_{\rm survey}}{df_l} \propto n_l(f_l)~{\cal N}(f_l),
\label{eq:dNsurv}
\end{equation}
where we highlight the fact that both the number density of lens stars and the number of events per disk crossing depend on $f_l$.  For comparison, consider the number of classic stellar microlensing events in the same survey, which has the same form as equation (\ref{eq:Nsurv}) but with $a_{\rm out} \to b_{\phi\star} = \phi R_{E\star}$ and ${\cal N} \to 1$ (since by definition the number of microlensing events when the source crosses a lensing region is unity).  Since $b_{\phi\star} \propto R_{E\star} \propto \sqrt{f_l(1-f_l)}$, we have
\begin{equation}
\frac{dN_{\rm survey,\star}}{df_l} \propto n_l(f_l) \sqrt{f_l(1-f_l)} .
\label{eq:dNstar}
\end{equation}

Finally, we can quantify the distribution of microlensing event durations.  While the Einstein crossing time is given by equation (\ref{eq:tein}), the actual event duration also has a factor of the boost: $t_{\rm lens} = 2\phi R_{\rm E}/v_\perp$.  This depends on mass (through both $R_{\rm E}$ and $\phi$), so it is useful to compute the weighted mean:
\begin{equation}
\begin{split}
  \langle t_{\rm lens} \rangle &= \frac{1}{\cal N} \int t_{\rm lens}~
    \frac{d{\cal N}}{dm}~dm, \\
  &= \frac{4 \sqrt{G\tilde{D}}}{v_\perp c}~
    \frac{\int \phi^2~m^{1-\alpha}~dm}{\int \phi~m^{1/2-\alpha}~dm}.
\end{split}
\label{eq:tavg}
\end{equation}
Note that
\begin{equation}
\frac{\langle t_{\rm lens} \rangle}{t_{\rm disk}} \approx \frac{2 \tau}{{\pi \cal N}}.
\label{eq:tlenstdisk}
\end{equation}

The limits of integration for the mass integrals are $\{m_{\rm min},m_{\rm L} \}$.  The minimum detectable mass $m_{\rm min}$ results from two conditions.  The first is that the Einstein radius of the planetesimal exceeds its physical radius, $r>R_{\rm E}$, which is satisfied for all but the lowest values of $m$ and $f_l$ considered (see Figure \ref{fig:minmass}). The second condition requires the planetesimal to be ``massive enough'' to microlens a source that has some finite projected size $R_\star$.  We quantify this condition by considering the ratio of the angular source size to the angular Einstein radius, which we denote by $\rho_\star$ (see \S\ref{subsect:finite_source}).  Collectively, the two conditions yield the minimum mass
\begin{equation}
m_{\rm min} = \mbox{max}\left\{\frac{9 c^6}{1024 \pi^2 \rho_p^2 G^3 \tilde{D}^3}, \frac{\epsilon \left(c f_l R_\star\right)^2}{4G \tilde{D}} \right\},
\label{eq:m_min}
\end{equation}
where
\begin{equation}
\epsilon = \left( \min_\delta\left\{1/\rho_\star \right\} \right)^2 < 1.
\end{equation}
The minimum value of $1/\rho_\star$ effectively quantifies the smallest planetesimal mass that can produce a fractional flux change $\delta \ge \delta_{\rm det}$, given the finite size of the source.  To be in the amplitude-limited regime, a third condition on $m_{\rm min}$ that involves $t_{\rm min}$ is applicable,
\begin{equation}
m_{\rm min} \ge \frac{\epsilon \left(v_\perp c t_{\rm min} \right)^2}{16 G \tilde{D}}.
\end{equation}
As discussed after equation (\ref{eq:duration-limited}), we do not explicitly include this condition in our calculations, but we do consider how it will affect our conclusions.

\subsection{Zeroth Order Estimates}
\label{subsect:fom_zeroth}

To make initial estimates, we require that the impact parameter be smaller than the Einstein radius (i.e., $\phi=1$), which amounts to setting the detection threshold to $\delta_{\rm det}=0.34$ for a small source.  We also require that the Einstein radius be larger than the projected size of the source (i.e., $\epsilon=1$).  When $\phi$ is independent of the planetesimal mass, equations (\ref{eq:size_dist}) and (\ref{eq:tau}) yield $d\tau/dm \propto m^{1-\alpha} \propto dM_{\rm disk}/dm$.  In other words, the optical depth is directly proportional to the total disk mass --- but insensitive to how the planetesimal masses are distributed --- above $m_{\rm min}$.  Since the total disk mass is dominated by the high-mass end of the planetesimal mass distribution, we have
\begin{equation}
\tau = \frac{4 G \tilde{D}}{\left( a c \right)^2} M_{\rm disk} \left[1 - \left(\frac{m_{\rm min}}{m_{\rm L}}\right)^{2-\alpha} \right] \approx \frac{4 G \tilde{D}}{\left( a c \right)^2} M_{\rm disk}.
\label{eq:tau0}
\end{equation}
If we fix the source and vary the distance to the lens, the distance $\tilde{D} = f_l(1-f_l) D_s$ peaks at $f_l = 0.5$.  The peak can be shifted slightly because $m_{\rm min}$ also depends on $f_l$, but our full calculation indicates that optical depth is still maximized near $f_l \approx 0.5$ (see \S\ref{subsect:fom_improved}).  The disk parameters in equation (\ref{eq:parameters}) yield $\tau \approx 5 \times 10^{-6}$ for $f_l = 0.5$.  Thus, even if there is a source behind a planetesimal disk, the instantaneous probability of disk microlensing is low.

The expected number of microlensing events during one disk crossing is
\begin{equation}
{\cal N} \approx \frac{4 \sqrt{G \tilde{D}}}{\pi a c} \left(\frac{4-2\alpha}{3-2\alpha}\right) \left(\frac{m^{3/2-\alpha}_{\rm L} - m^{3/2-\alpha}_{\rm min}}{m^{2-\alpha}_{\rm L}}\right) M_{\rm disk}.
\label{eq:events0}
\end{equation}
Here the fact that $m_{\rm min}$ depends on $f_l$ becomes important; it shifts the peak in ${\cal N}$ to values somewhat lower than $f_l = 0.5$, and also determines the slope of the ${\cal N}$ versus $f_l$ curve.  In particular, $m_{\rm min} \propto f_l/(1-f_l)$ while $\tilde{D} \propto f_l(1-f_l)$, so for small $f_l$ we find
\begin{equation}
{\cal N} \propto f_l^{2-\alpha} = f_l^{1/6}
\end{equation}
for $\alpha=11/6$.  With $D_s = 8$ kpc, $R_\star = R_\sun$ and $m_{\rm L} = 2 M_\earth$, we find that ${\cal N}$ peaks when $f_l \approx 0.09$ ($m_{\rm min} \approx 0.01 M_\earth$), with a value of $\max\{{\cal N}\} \approx 4 \times 10^{-3}$.  This is still in the fiducial amplitude-limited regime (cf. equation \ref{eq:duration-limited}), so the scaling with $f_l$ is accurate.  The scaling differs in the duration-limited regime, but this is of less interest to us in the present analysis.

Using equation (\ref{eq:tlenstdisk}), the mean microlensing duration is
\begin{equation}
\langle t_{\rm lens} \rangle \approx \frac{4 \sqrt{G\tilde{D}}}{v_\perp c}~ \left(\frac{3-2\alpha}{4-2\alpha}\right) \frac{m^{2-\alpha}_{\rm L}}{m^{3/2-\alpha}_{\rm L} - m^{3/2-\alpha}_{\rm min}}.
\end{equation}
Adopting the same parameter values as in the preceding paragraph, we have $\langle t_{\rm lens} \rangle \approx 2$ hours when $f_l \approx 0.09$.

\subsection{Finite Source Effects}
\label{subsect:finite_source}

The simple estimates we just obtained should underestimate the occurrence of planetesimal disk microlensing for two reasons.  Firstly, the assumed detection threshold of $\delta_{\rm det}=0.34$ was very generous.  Secondly, we explicitly neglected planetesimals whose Einstein radius is smaller than the projected size of the source star.  If we account for finite source microlensing, we can consider the possibility that smaller planetesimals may contribute to the overall signal (di Stefano \& Scalzo 1999a).

In this analysis, it is useful to introduce angular variables:
\begin{equation}
\theta_{\rm E} = \frac{R_{\rm E}}{D_l},
\quad
\theta_b = \frac{b}{D_l},
\quad
\theta_\star = \frac{R_\star}{D_s},
\end{equation}
where $b$ is the impact parameter.  The source size is important in relation to the Einstein radius, so we consider the ratio
\begin{equation}
\rho_\star \equiv \frac{\theta_\star}{\theta_{\rm E}}.
\end{equation}
If we define
\begin{equation}
u_0 \equiv \frac{b}{R_{\rm E}} = \frac{\theta_b}{\theta_{\rm E}},
\end{equation}
then the lensing magnification ${\cal A}(u_0,\rho_\star)$ formally involves a two-dimensional integral, but Heyrovsky \& Loeb (1997) and Lee et al. (2009) give a useful algebraic approximation (see Appendix), allowing computations to be significantly accelerated.

\begin{figure}
\begin{center}
\includegraphics[width=4in]{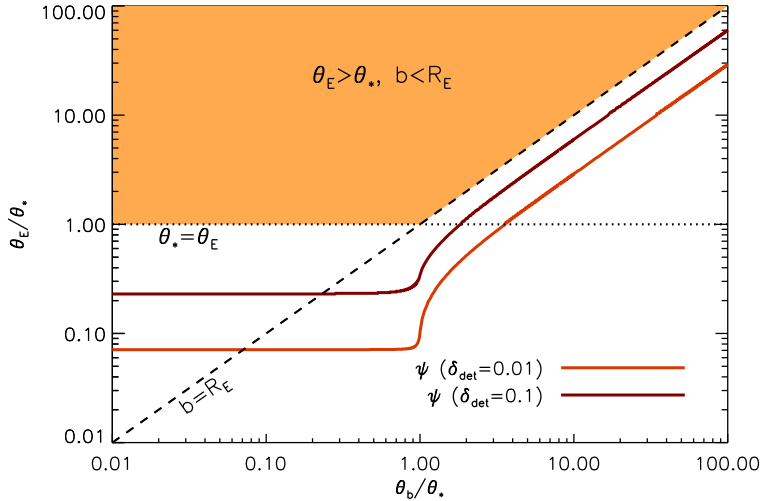}
\end{center}
\caption{In the plane of impact parameter and Einstein radius (both in angular units, normalized by the size of the source star), the shaded region denotes the phase space used for zeroth order estimates of the microlensing figures of merit (\S\ref{subsect:fom_zeroth}).  The $\psi$ curves show how the allowed phase space is enlarged when we consider different detection thresholds ($\delta_{\rm det}$) and account for finite source effects (\S\ref{subsect:finite_source}).}
\label{fig:finite_source}
\end{figure}

We evaluate ${\cal A}$ in the plane of $\theta_{\rm E}/\theta_s$ and $\theta_b/\theta_\star$.  Some portion of this phase space satisifies $\delta \ge \delta_{\rm det}$; this region is bounded by the curve,
\begin{equation}
\psi = \psi\left(m, f_l, b=b_\phi \right),
\end{equation}
which can be inverted to solve for $\phi$ and $\epsilon$.  Examples of $\psi$ curves for $\delta_{\rm det} = 0.01$ and 0.1 are shown in Figure~\ref{fig:finite_source}.  For comparison, the shaded region in the figure shows the region of phase space used for the zeroth order estimates in \S\ref{subsect:fom_zeroth} (specifically, the region bounded by the curves $\theta_\star = \theta_{\rm E}$ and $b = R_{\rm E}$).  Allowing a more sensitive detection threshold and accounting for finite source microlensing clearly increases the region of phase space that can contribute to the microlensing signal.

\begin{figure}
\begin{center}
\includegraphics[width=4in]{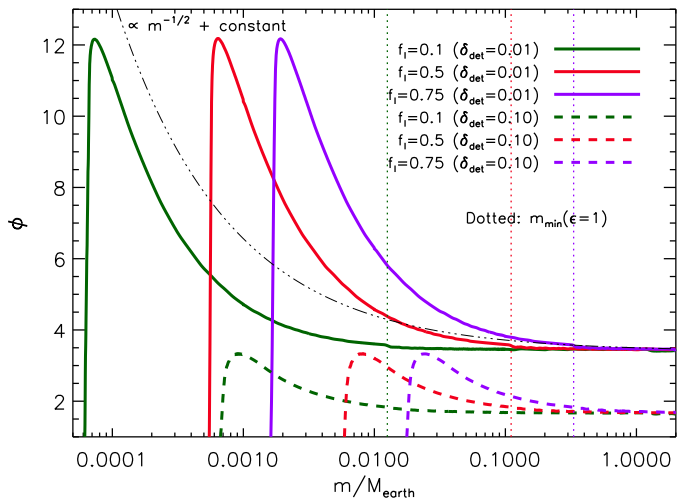}
\end{center}
\caption{Boost factor $\phi$ as a function of planetesimal mass $m$, for different values of the lens/source distance ratio ($f_l$) and the detection threshold ($\delta_{\rm det}$). The dotted vertical lines indicate the minimum detectable mass if the Einstein radius is required to be larger than the projected source size ($\epsilon=1$).  The dot-dot-dot-dash curve shows the heuristic $\phi$ curve from equation (\ref{eq:phisimple}).}
\label{fig:phi}
\end{figure}

\begin{figure}
\begin{center}
\includegraphics[width=4in]{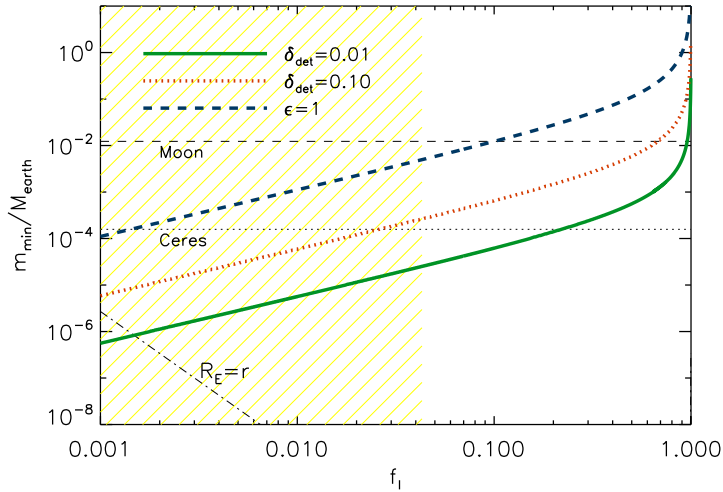}
\end{center}
\caption{Minimum detectable mass $m_{\rm min}$ as a function of the lens/source distance ratio $f_l$, for different values of the detection threshold $\delta_{\rm det}$.  The shaded region shows the nominal duration-limited regime (see equation [\ref{eq:duration-limited}]).  The thin dot-dash line shows the point at which the Einstein radius is the same as the size of planetesimal ($R_{\rm E}=r$); for the detection thresholds we consider, the point lens approximation breaks down only at $f_l \lesssim 10^{-3}$.  We note that when $R_{\rm E} \approx r$ there can be simultaneous microlensing and occultation, which probe the physical size of the lens (Bromley 1996; Agol 2002).  The horizontal dashed and dotted lines indicate the masses of the Moon and Ceres, respectively.}
\label{fig:minmass}
\end{figure}

To illustrate the enhancement, we plot the boost factor $\phi$ as a function of the planetesimal mass in Figure~\ref{fig:phi}.\footnote{This figure is reminiscent of the bottom panel of Figure 6 of Han et al. (2005), where they adopted $f_l=0.75$ with $D_s = 8$ kpc.}  The shape of the $\phi(m)$ curve is not very sensitive to the value of $f_l$, which can be understood heuristically --- the maximum impact parameter roughly corresponds to a situation in which the near edge of the source star is some distance ${\cal C} R_{\rm E}$ away from the lens, where ${\cal C}$ is a constant that depends on the detection threshold $\delta_{\rm det}$:
\begin{equation}
  b_\phi \sim f_l R_\star + {\cal C} R_{\rm E},
\end{equation}
where $f_l R_\star$ is the size of the source star projected into the lens plane.  This is equivalent to the boost factor being
\begin{equation}
  \phi \sim \frac{f_l R_\star}{R_{\rm E}} + {\cal C}.
\label{eq:phisimple}
\end{equation}
The constant ${\cal C}$ is described by equation (4) of di Stefano \& Scalzo (1999a); we checked that our values of ${\cal C}$ in Figure \ref{fig:phi} are consistent with their formula.  Since $R_{\rm E} \propto m^{1/2}$, we expect that $\phi$ is approximately constant at large $m$, then rises as $m$ decreases, down to a minimum threshold mass. This heuristic shape is shown in Figure~\ref{fig:phi}. We see that this general argument explains the overall shape of the $\phi$ curves, although it is too simple to capture the full complexity near the minimum mass.

Finite source effects dictate that the boost factor can be as high as $\sim$3 or 12, for a detection threshold of $\delta_{\rm det}=0.1$ or 0.01, respectively (i.e., a ten or one percent perturbation).  A second result is that the minimum mass that contributes to microlensing can be smaller than one naively estimates by requiring the Einstein radius to be larger than the projected source size.  For $\delta_{\rm det}=0.1$ or 0.01, $m_{\rm min}$ is reduced (relative to the naive estimate) by a factor of $1/\epsilon \sim 20$ or 200, respectively.  For example, if the disk is located at $f_l=0.5$ and we consider $\delta_{\rm det}=0.1$, we have $m_{\rm min} \approx 5 \times 10^{-3} M_\earth$ instead of about 0.1 $M_\earth$.  More generally, Figure~\ref{fig:minmass} shows $m_{\rm min}$ as a function of $f_l$ for different detection thresholds, demonstrating that \emph{disk microlensing can, in principle, probe planetesimals down to several orders of magnitude below Earth mass.}

It is important to note that Han et al. (2005) conclude it will be difficult to probe masses below $\sim 0.02~M_\earth$ (at $f_l = 0.75$) with microlensing.  One key difference is that Han et al.\ (2005) consider a stringent limit for their signal-to-noise ratio of $S/N = \sqrt{1000}$.  For comparison, with $f_l = 0.75$ and $\delta_{\rm det} = 0.1$ we find $m_{\rm min} \sim 0.02~M_\oplus$ (Figure \ref{fig:phi}).  Using equation (17) of Han et al. (2005), we obtain $S/N \sim 29$ for our value of $m_{\rm min}$, assuming only one detection point per crossing.  Plugging in our estimate for $S/N$ into their equation (18) then yields a minimum mass that is consistent with our computed value.  A less stringent criterion for the $S/N$ threshold (e.g., $S/N \approx 8$ by Gaudi et al. 2002) will lead to a different value for $m_{\rm min}$.  If we consider $\delta_{\rm det} = 0.01$, $m_{\rm min} \sim 2 \times 10^{-3}~M_\oplus$ (Figure \ref{fig:phi}) and retain all of the numbers in the preceding estimate, we obtain $S/N \sim 6$ instead and are consistent with equation (18) of Han et al. (2005) to within a factor of 2.  Therefore, we believe our calculations are consistent and the issue is more one of deciding what constitutes a detectable event (see \S\ref{subsect:detectability} for more discussion).

\subsection{Improved Estimates of the Figures of Merit}
\label{subsect:fom_improved}

\begin{figure}
\begin{center}
\includegraphics[width=4in]{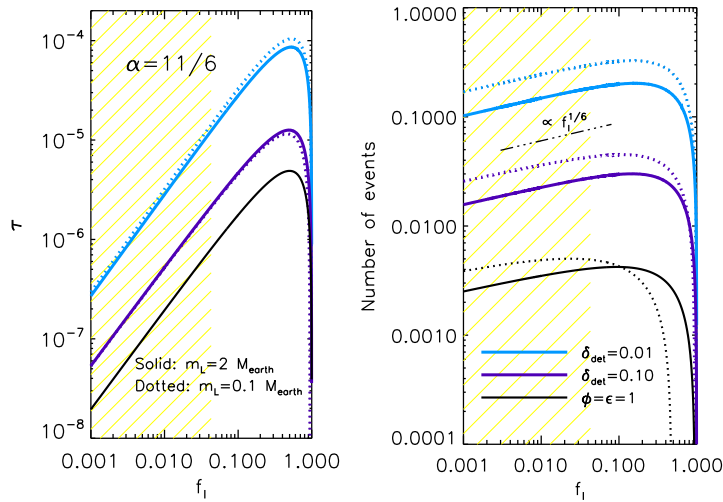}
\end{center}
\caption{Figures of merit for planetesimal microlensing as a function of the lens/source distance ratio $f_l$: optical depth $\tau$ (left); and expected number of events per disk crossing ${\cal N}$ (right).  The various curves are for different values of the largest planetesimal mass $m_{\rm L}$ and the detection threshold $\delta_{\rm det}$.  (The black curves correspond to the zeroth order estimates in \S \ref{subsect:fom_zeroth}.)  The curves are reliable in the amplitude-limited regime, but are overestimates in the duration-limited regime (shaded; see equation [\ref{eq:duration-limited}]).}
\label{fig:fom}
\end{figure}

We can now obtain better estimates of the optical depth and number of events per planetesimal disk crossing, taking into account finite source effects to properly treat the full range of allowed lens masses.  Figure~\ref{fig:fom} shows examples of $\tau$ and ${\cal N}$ as a function of $f_l$ for different detection thresholds as well as two values of the largest planetesimal mass, $m_{\rm L}=0.1$ and $2~M_\earth$; the latter value is the maximum allowed mass for our planetesimal disk configuration (see \S\ref{subsect:dynamics}).  For illustration, we continue to set $\alpha=11/6$.

The optical depth peaks when the lens is approximately halfway to the source ($f_l \approx 0.5$), and scales as $\tau \propto f_l$ when the lens is near the observer (i.e., $f_l$ is small).  Reducing $m_{\rm L}$ from $2~M_\earth$ to $0.1~M_\earth$ means the disk mass is distributed among a larger number of intermediate-mass planetesimals.  This has little effect on the zeroth order estimate for the optical depth (the black curve in the left panel of Figure \ref{fig:fom}; cf.\ equation [\ref{eq:tau0}]), but it does cause a modest increase in the more realistic optical depth estimates --- at least for small $f_l$ values --- because the boost factor enhances the effects of intermediate-mass planetesimals.  For $\delta_{\rm det}=0.1$, the peak and drop-off in the $\tau$ curve occur at smaller values of $f_l$ for $m_{\rm L}=0.1~M_\earth$ than for $m_{\rm L}=2~M_\earth$, simply because the planetesimals become too small to produce $\delta \ge \delta_{\rm det}$. On average, the more realistic optical depth is boosted (relative to the zeroth order estimate) by a factor $\sim$3 or 20 for a detection threshold of $\delta_{\rm det} = 0.1$ or 0.01, respectively.  Even so, the optical depth remains small.

The number of events per disk crossing is considerably more sensitive to the value of $m_{\rm L}$, because this figure of merit is more sensitive to low-mass planetesimals.  For $m_{\rm L} = 2~M_\earth$, the realistic estimate for ${\cal N}$ is a factor $\sim$9 or 60 larger than the zeroth order estimate, for $\delta_{\rm det} = 0.1$ or 0.01, respectively; and for $m_{\rm L} = 0.1~M_\earth$ the enhancement factors are $\sim$19 or 140.  Nevertheless, we still find ${\cal N} < 1$ for typical parameters, which means we are unlikely to see multiple microlensing events from a single disk (for our fiducial disk parameters).

\begin{figure}
\begin{center}
\includegraphics[width=4in]{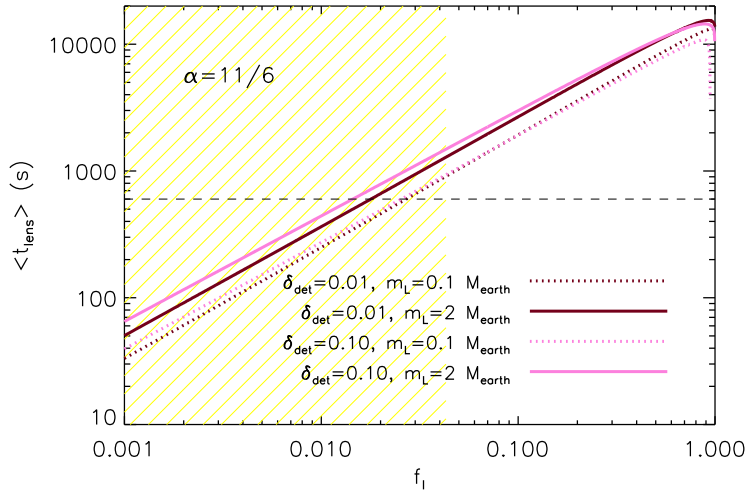}
\end{center}
\caption{Mean microlensing event duration (cf.\ equation [\ref{eq:tavg}]), as a function of the lens/source distance ratio $f_l$.  The horizontal dashed line indicates $\langle t_{\rm lens} \rangle = 10$ min, which is about the best cadence that current microlensing surveys are capable of.  The curves are reliable in the amplitude-limited regime, but are {\it underestimates} in the duration-limited regime (since events that are too short to be detected are excluded).  We have assumed $v_\perp = 100$ km s$^{-1}$, but the duration can be trivially rescaled for other choices.}
\label{fig:tavg}
\end{figure}

The mean event duration decreases with $f_l$, as shown in Figure~\ref{fig:tavg}.  This actually reconciles the low optical depth with the modest total number of events: any single observation is unlikely to catch a short, transient event, but a thorough monitoring program will be able to find it.  

\begin{figure}
\begin{center}
\includegraphics[width=4in]{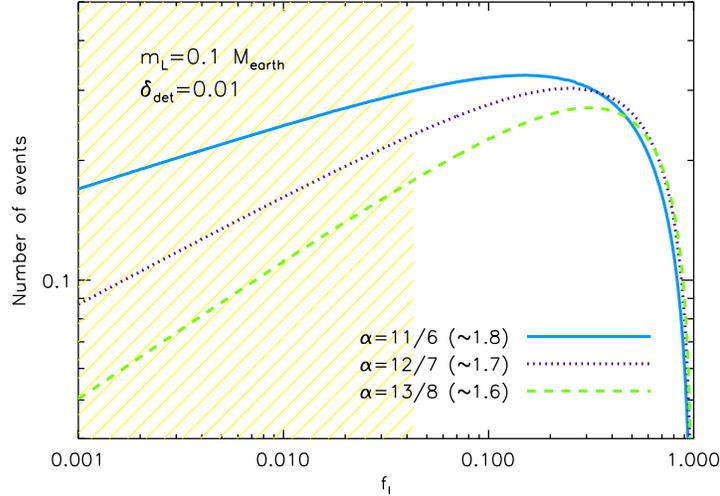}
\end{center}
\caption{Expected number of microlensing events per planetesimal disk crossing for various values of the planetesimal mass distribution index $\alpha$.  We adopt $m_{\rm L} = 0.1$ $M_\earth$ and $\delta_{\rm det}=0.01$ for illustration.  The shaded region again shows the nominal duration-limited regime.}
\label{fig:fom_otheralphas}
\end{figure}

The weak dependence of ${\cal N}$ on $f_l$ stems from having a steep mass distribution of planetesimals.  If all of the planetesimals had the same mass, we obtain the scaling ${\cal N} \propto f^{1/2}_l$ (for small $f_l$).  Instead, with a planetesimal mass distribution $dN/dm \propto m^{-\alpha}$ we find ${\cal N} \propto f_l^{2-\alpha}$ (again for small $f_l$).  The dependence on $\alpha$ is illustrated in Figure~\ref{fig:fom_otheralphas}, which shows ${\cal N}$ versus $f_l$ for the other $\alpha$ values described in \S\ref{subsect:popn}.  We see that for $\alpha < 11/6$, ${\cal N}$ decreases by factors $\sim 2$ to 3, but still remains non-negligible.  All of the scalings discussed here apply to the amplitude-limited regime; the scalings with $f_l$ are steeper in the duration-limited regime, because at low $f_l$ the events created by the lowest mass planetesimals become too short to be reliably detectable.

\subsection{Distribution of Distance Ratios for Detected Events}
\label{subsect:fdist}

Having understood how the figures of merit depend on $f_l$, we can now make simple estimates of the distribution of $f_l$ values that may be found in a microlensing survey.  The key difference from the preceding analysis is that we need to combine the raw figures of merit with a reasonable estimate for the spatial distribution of lens stars.  

We use equations (\ref{eq:dNsurv}) and (\ref{eq:dNstar}) to interpret the probability density of detecting microlensing events at a given $f_l$ value as
\begin{equation}
\begin{split}
p_{\rm disk}\left(f_l\right) \propto \frac{dN_{\rm survey}}{df_l},\\
p_\star\left(f_l\right) \propto \frac{dN_{\rm survey,\star}}{df_l}.\\
\end{split}
\end{equation}
The proportionality constant in each case serves to normalize the probability density distribution such that it integrates (over $f_l$) to unity.

The source stars are assumed to be in the Galactic bulge.  For the lens stars residing in the disk of the Galaxy, the density distribution can be approximated with a simple exponential disk model,
\begin{equation}
n_l \propto \exp{\left(-\frac{R_g}{H}\right)},
\end{equation}
where $R_g = (1-f_l)D_s$ is the distance from the Galactic center and $H = 3$ kpc is our adopted value for the radial scale length (sometimes termed ``radial scale height'') of the Galactic disk (Ojha 2001).  For the lens stars residing in the Galactic bulge, the density distribution is
\begin{equation}
n_l \propto R_g^{-s},
\end{equation}
where the power law index is either $s=1.75$ (Binney et al. 1991) or $s=1.85$ (Kent 1992).

\begin{figure}
\begin{center}
\includegraphics[width=4in]{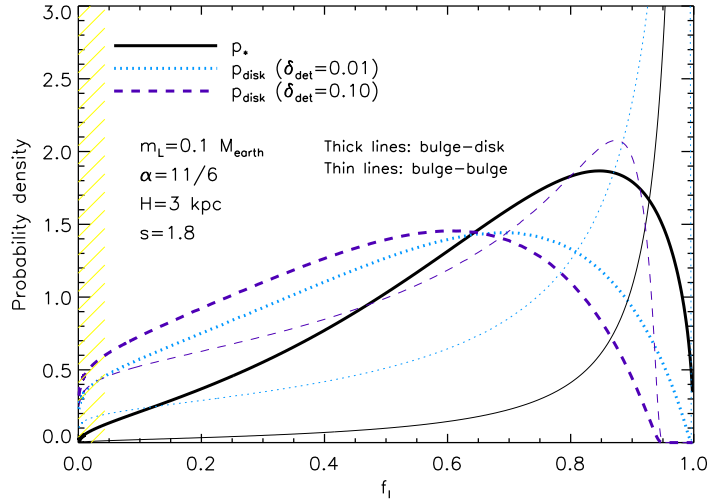}
\end{center}
\caption{Probability density for the lens/source distance ratio $f_l$ for both classic ($p_\star$) and planetesimal disk microlensing ($p_{\rm disk}$) events.  The source stars are assumed to be in the Galactic bulge, while we consider the lens stars to be both in the Galactic bulge (thin lines) and disk (thick lines).  The nominal duration-limited regime is again shaded.}
\label{fig:fdist}
\end{figure}

Our estimates for $p_{\rm disk}$ and $p_\star$ are shown in Figure \ref{fig:fdist}, where we assume $m_{\rm L} = 0.1~M_\oplus$ and $\alpha=11/6$.  For bulge-disk microlensing, the classic events peak at $f_l \approx 0.85$, while the planetesimal disk events peak at $f_l \approx 0.69$ ($\delta_{\rm det} = 0.01$) and $f_l \approx 0.62$ ($\delta_{\rm det} = 0.1$).  It is interesting to see that the $f_l$ distribution for a survey with the poorer detection threshold ($\delta_{\rm ret}=0.1$) is shifted to lower values compared to a survey with the better detection threshold ($\delta_{\rm ret}=0.01$).  The reason is that for $\delta_{\rm det} = 0.1$ the minimum detectable mass can exceed the maximum planetesimal mass, i.e., $m_{\rm min} > m_{\rm L}$, for high values of $f_l$, which causes the microlensing signal to vanish about $f_l \approx 0.95$.  For bulge-bulge microlensing, all of the events peak at $f_l \ge 0.87$ for $1.75 \le s \le 1.85$; we adopt $s=1.8$ in Figure \ref{fig:fdist} for illustration.  In all cases, planetesimal disk microlensing events always peak at lower $f_l$ values compared to classic microlensing events.

We note that Kiraga \& Paczy\'{n}ski (1994) estimate bulge-bulge microlensing events to be about 5 times more likely than bulge-disk ones.

\section{Discussion}
\label{sect:discussion}

\subsection{Detectability}
\label{subsect:detectability}

In this paper, we have adopted the simple criterion that a planetesimal disk microlensing event occurs when a planetesimal amplifies background starlight above some specified threshold for longer than some minimum duration.  While this criterion does not account for all of the subtleties involved in a realistic observational campaign, we believe it is useful for elucidating the general properties of planetesimal disk microlensing.  Conceptually, our main conclusions are:
\begin{enumerate}

\item It is conceivable to detect planetesimals in the range of the mass of the Moon or even Ceres --- and possibly below --- although pushing to lower masses becomes increasingly challenging.

\item Planetesimal disk microlensing events are likely to be shifted to lower values of $f_l$ than classic stellar microlensing events, because decreasing $f_l$ lowers the minimum planetesimal mass that contributes to the microlensing signal, thereby increasing the number of available deflectors.  
\end{enumerate}
Quantitatively, we estimate that if it is possible to detect microlensing events in which a solar-type source star is amplified by at least 10\%, it will only take a few tens of disk crossings to discover one planetesimal microlensing event; and if it is possible to detect a 1\% brightening, it will only take a handful of planetesimal disk crossings (for our fiducial disk parameters).

It is still important to consider some practical aspects of microlensing, especially with regards to understanding how they will affect our conclusions.  By our adopted criterion, a single deviant data point in a light curve is regarded as a detection.  A cautionary example comes from the study of Sahu et al. (2001), who detected 6 temporally unresolved spikes in their data while scrutinizing the globular cluster M22, and initially interpreted them as microlensing events caused by free-floating planets with masses $\sim 0.25 M_{\rm J}$, where $M_{\rm J}$ is the mass of Jupiter.  Gaudi (2002) argued on dynamical grounds that it is implausible for free-floating planets to exist in the core of M22, and that any such planets in the halo will not be numerous enough to explain the observed microlensing optical depth.  Eventually, Sahu et al. (2002) discovered that the ``microlensing spikes'' were caused by cosmic rays.

The statistical significance of a microlensing event ultimately depends not just on the maximum fractional flux deviation but on more sophisticated quantities such as the total signal-to-noise ratio ($S/N$; Han et al.\ 2005) or the $\chi^2$ difference between the goodness of fit for model light curves with and without planets/planetesimals (Gaudi \& Sackett 2000).  There is no solid consensus on what should constitute a detection and various thresholds have been adopted in the literature: $S/N = \sqrt{1000} \approx 32$ (Han et al.\ 2005), $S/N =\sqrt{60} \approx 8$ (Gaudi et al.\ 2002) or $S/N=12.5$ (Bennett et al.\ 2004), and $\Delta \chi^2 \ge 100$ or 225 (Gaudi \& Sackett 2000).   For our example in \S\ref{subsect:finite_source}, the detection thresholds of $\delta_{\rm det} = 0.01$ and 0.1 are equivalent to $S/N \sim 6$ and 29, respectively.

If the planetesimals have a steep mass distribution function, then the number of events per disk crossing remains non-negligible even for nearby planetesimal disks ($f_l \ll 1$).  In principle, the closer the disk, the farther down the planetesimal mass function we can probe.  However, events with small $f_l$ are expected to be relatively rare given realistic lens and source populations (see Figure \ref{fig:fdist}).  Also, their short durations will make them more difficult to detect, since at fixed cadence an event due to a nearby planetesimal event will have fewer deviant light curve points and thus a lower total $S/N$ than a corresponding event due to a more distant planetesimal disk.  Finally, for events with $f_l \ll 1$ the microlensing signal will be more diluted by light from the lens star.  Consider our illustration in Figure \ref{fig:cartoon}, where $f_l = 0.1$.  Even without taking dust extinction into account, the lens star is $\sim$100 times brighter than the source star.  Since the lens and source cannot be resolved during the microlensing event, a deviation of 10\% in the source flux will become merely a $\sim$0.1\% deviation in the total observed flux.  This issue becomes yet more severe for systems with smaller $f_l$ values.  These issues seem to make it very challenging to detect microlensing events from very nearby planetesimal disks.  Nevertheless, we recall that planetesimal disk microlensing events are still expected to have somewhat lower values of $f_l$ than stellar microlensing events (Figure \ref{fig:fdist}).

\subsection{Observational Strategies}
\label{subsect:strategies}

Traditional microlensing surveys usually target a field of source stars and wait for lenses to pass in front.  Next-generation campaigns will use wide-field cameras mounted on 1--2m telescopes to monitor $\sim 10$ deg$^2$ fields at cadences of 10 minutes (Bennett et al.\ 2009; Gaudi et al.\ 2009).  Such cadences should be able to detect microlensing events associated with all but the closest planetesimal disks.  At this point it is difficult to forecast in detail the number of planetesimal disk microlensing events that will be detected due to uncertainties in the abundance of planetesimal disks.

In the wide-separation limit considered, a planetesimal disk microlensing event may or may not be accompanied by an event associated with the parent star --- isolated, short-duration events may occur (as has been discussed for planets; Di Stefano \& Scalzo 1999a,b) and will be quite interesting.  For an isolated planetesimal event, it may be difficult to determine the actual planetesimal mass because the Einstein radius crossing time is a degenerate combination of the lens mass, the lens and source distances, and the transverse relative velocity (Gaudi 2002; Han et al.\ 2005).\footnote{When $\theta_{\rm E} \sim \theta_\star$, the resulting magnification pattern has structure on the scale of the projected source size, which allows $\rho_\star$ to be measured, thus breaking some of the degeneracy.}  For an event that shows effects from both the planetesimal and the parent star, the light curve alone will strongly constrain the mass \emph{ratio} between the two lenses.  Working with mass ratios will already enable interesting planetesimal disk science.

A plausible, alternate strategy will be to target known debris disks and wait for source stars to move behind them.  As we have seen, the microlensing signal is appreciable even for disks that are relatively nearby.  The practical challenge is to locate a disk with a suitable star behind it.  Typical debris disk searches tend to avoid crowded stellar fields, because of the difficulties with the point spread function subtraction that impede the measurement of an IR excess (A. Moro-Mart\'{i}n 2009, private communication).  In this regard it is unclear whether existing debris disk samples offer good candidates for microlensing follow-up.  It is worthwhile to consider whether there are observational strategies that can combine debris disk observations with microlensing to reap the benefits of both.

While discovering planetesimal disk microlensing events will obviously be exciting, even the non-detection of planetesimals in microlensing lightcurves will set interesting upper limits on their masses that will be useful to models of debris disks.  Wyatt \& Dent (2002) proposed that dust clumps embedded in the debris disk of Fomalhaut are the result of collisions between planetesimals that may be as large as $\sim 1000$ km in size ($\sim 0.2$ lunar masses or $\sim 13$ times the mass of Ceres).  Such estimates hinge on uncertain extrapolations based on 450 $\mu$m and 850 $\mu$m observations of $\sim 7$ $\mu$m and $\sim 0.2$ m objects.  As mentioned in \S\ref{sect:intro}, determining the size of the largest planetesimal remains an open question in debris disk studies.  One can begin to address this question by examining the microlensing lightcurves of hundreds of source (dwarf) stars and searching for statistically significant residuals.  In principle, if the distances to the source and parent stars, the relative velocity between them, and the planetesimal disk geometry are known, one can infer the maximum planetesimal mass detectable for a given magnification threshold.  Non-detections will also set constraints on the size of the planetesimal disk.  Surveying a large number of stars for planetesimal microlensing events will shed light on the frequency of planetesimal disks with or without planets, thus constraining models of planet formation.

\subsection{Future Work}

In this paper, we have considered large disks in which the planetesimals lie ``far'' from the parent star (relative to the stellar Einstein radius), and focused on events in which the source passes close to a planetesimal.  There are two interesting ways to extend our analysis.  One way is to account for tidal shear, which may be created not only by the parent star but also by other planetesimals; this will allow us to analyze small disks.  The second approach is to consider a scenario in which the source passes so close to the parent star that the light curve is affected by caustics created by the planetesimals.  This is a direct analog of high-magnification microlensing events that is used to detect planets (Wambsganss 1997; Griest \& Safizadeh 1998; Gaudi et al.\ 1998; Gould 2008), but generalized from the case of one or a few massive planets to many planetesimals.  The caustics are sensitive to the full population of planetesimals, so a high-magnification microlensing event will probe the entire planetesimal disk simultaneously.

A shortcoming of our idealized analysis is the exclusion of blending effects, in particular the assumption in equation (\ref{eq:fobs}) that $F_{\rm lens}=0$.  In practice, the lens and source stars are located within each other's Einstein radius during a microlensing event, and within the resolution limit of current ground- and space-based telescopes.  We intend to explore this issue in a future paper.

Clearly there is much fertile ground for further work on both the formal and practical aspects of planetesimal disk microlensing.  We believe the possibility of obtaining a new way to analyze planetesimal disks will make such studies interesting and exciting.

\scriptsize
\acknowledgements
We acknowledge generous financial, computational, logistical and secretarial support from the Institute for Advanced Study (IAS).  KH is the Frank \& Peggy Taplin Member at the IAS, and also receives support from NASA grant NNX08AH83G and NSF grant AST-0807444.  CRK receives support from NSF through grant AST-0747311.  CRK thanks the astrophysics group at the IAS for its hospitality during an extended visit, when this project was conceived.  We are indebted to Scott Tremaine for invaluable suggestions and guidance.  We are also grateful to Scott Gaudi, Subo Dong, Margaret Pan, Charles Beichman, Takahiro Sumi, Christine Chen, Joachim Wambsganss, Doug Lin, Amaya Moro-Mart\'{i}n and Zheng Zheng for useful conversations, many of which occurred at the {\it 2nd Subaru International Conference} in Kona, Hawaii.  Finally, we thank the anonymous referee for constructive criticism that improved the quality of the paper.
\normalsize


\appendix

\section{Functional Form of Magnification with Finite Source}
\label{append:lee09}

The magnification of a finite source by a point lens has been studied by Heyrovsky \& Loeb (1997) and Lee et al. (2009).  Equation (7) from Lee et al. (2009) gives a useful approximation for the case of a uniform, circular source:
\begin{equation}
{\cal A}\left(u_0, \rho_\star\right) \approx 
\begin{cases}
\frac{1}{2 \rho_\star^2 n_{\rm res}} \left[{\cal F}_0 + \sum^{2n_{\rm res} - 1}_{k=1} {\cal F}\left(\frac{k\pi}{2n_{\rm res}} \right) \right] &,\ u_0 \le \rho_\star,\\
\frac{\Theta_{\rm crit}}{\pi \rho_\star^2 n_{\rm res}} \left[{\cal F}_0 + \sum^{n_{\rm res} - 1}_{k=1} {\cal F}\left(\frac{k \Theta_{\rm crit}}{n_{\rm res}} \right) \right] &,\ u_0 > \rho_\star,\\
\end{cases}
\end{equation}
where we have defined
\begin{equation}
\begin{split}
\Theta_{\rm crit} &\equiv \arcsin{\left(\rho_\star/u_0\right)},\\
{\cal F}_0 &\equiv \frac{1}{2}\left[ \left(u_0+\rho_\star\right) \sqrt{\left(u_0+\rho_\star\right)^2 + 4} - \left(u_0-\rho_\star\right) \sqrt{\left(u_0-\rho_\star\right)^2 + 4} \right],\\
{\cal F}\left(\Theta\right) &\equiv u_2 \sqrt{u^2_2 + 4} - u_1 \sqrt{u^2_1 + 4}.\\
\end{split}
\end{equation}
The quantities $u_1$ and $u_2$ are given by
\begin{equation}
u_1 = 
\begin{cases}
u_0 \cos{\Theta} - \sqrt{\rho_\star^2 - u_0^2 \sin^2{\Theta}} &,\ u_0 > \rho_\star \mbox{ and } \Theta \le \Theta_{\rm crit},\\
0 &,\ \mbox{otherwise},\\
\end{cases}
\end{equation}
and
\begin{equation}
u_2 = 
\begin{cases}
u_0 \cos{\Theta} + \sqrt{\rho_\star^2 - u_0^2 \sin^2{\Theta}} &,\ u_0 \le \rho_\star \mbox{ or } \left\{u_0 > \zeta \mbox{ and } \Theta \le \Theta_{\rm crit}\right\},\\
0 &,\ \mbox{otherwise}.\\
\end{cases}
\end{equation}
The accuracy of the formula increases with the resolution, $n_{\rm res}$.

More generally, Heyrovsky \& Loeb (1997) give formulae for the microlensing of elliptical sources with non-uniform surface brightness profiles.


\end{document}